\documentclass[pra,twocolumn,showpacs]{revtex4}
\usepackage{amssymb}

\usepackage{amsmath}
\usepackage{braket}             
\usepackage{hyperref}
\usepackage{graphicx}
\usepackage{xcolor}

\begin{document}

\title{Berezinskii-Kosterlitz-Thouless transition in two dimensional random-bond XY model on a square lattice}
\author{Yi-Bo Deng and Qiang Gu\footnote[2]{Corresponding author: qgu@ustb.edu.cn}}
\affiliation{Department of Physics, University of Science and
Technology Beijing, Beijing 100083, China}
\date{\today}


\begin{abstract}
We perform Monte Carlo simulations to
study the two dimensional random-bond XY model on a square lattice.
Two kinds of bond randomness with the coupling coefficient obeying
the Gaussian or uniform distribution are discussed. It is shown that
the two kinds of disorder lead to similar thermodynamic behaviors if
their variances take the same value. This result implies that the
variance can be chosen as a characteristic parameter to evaluate the
strength of the randomness. In addition, the
Berezinskii-Kosterlitz-Thouless transition temperature decreases as
the variance increases and the transition can even be destroyed as
long as the disorder is strong enough.
\end{abstract}

\pacs{75.50.Lk, 75.40.Mg, 05.70.Fh, 05.70.Jk }

\maketitle

The Berezinskii-Kosterlitz-Thouless (BKT) transition in two
dimensional system has been extensively studied for decades since
the discovery of the exotic quasi-long-range order formed by the
binding of vortex-antivortex pairs.\cite{1,2} The simplest model
to demonstrate the BKT transition is the so-called 2-Dimensional
(2D) XY model and the Hamiltonian takes the form
\begin{align}
H  = -\sum_{\braket{ij}}J_{ij}\textbf{\emph{S}}_i\cdot\textbf{\emph{S}}_j =
-\sum_{\braket{ij}}J_{ij}\cos\theta_{ij},
\end{align}
where $\textbf{\emph{S}}_i = (S_i^x, S_i^y) = (\cos\theta_i, \sin\theta_i)$
donates spin at site $i$. $J_{ij}$ is the coupling coefficient of the
two nearest-neighboring sites, $i$ and $j$, and
$\theta_{ij}=\theta_i-\theta_j$ is the phase difference between the
two sites. Typically, the transition in the 2D XY model is
characterized by low temperature power-law decay of a two-point
correlation function which gives rise to divergent susceptibility,
\cite{3} measurable finite-size-induced magnetization with
universal magnetic exponent\cite{4} and a discontinuous jump to
zero of the helicity modulus.\cite{5} In experiments, the BKT
transition has been confirmed in various real systems such as $^4$He
films,\cite{6} Josephson-junction arrays\cite{6} and
planar lattice of Bose-Einstein condensates.\cite{8}

Due to the presence of defects and distortions, real systems are
always imperfect and usually subject to certain disorder effects.
Therefore it is of interest to study how the imperfection affects on
the BKT transition. For the 2D XY model, the imperfection can be
demonstrated by two parameters, $J_{ij}$ and $\theta_{ij}$. $J_{ij}$
(or $\theta_{ij}$) may be governed by a random distribution
$P(J_{ij})$ (or $P(\theta_{ij})$). It means that $J_{ij}$ (or
$\theta_{ij}$) takes the values subject to a probability
distribution, which is called the bond randomness (or phase
randomness).

Rubinstein, Shraiman, and Nelson studied 2D XY ferromagnets with
random Dzyaloshinskii-Moriya interactions and derived a model with
both $J_{ij}$ and $\theta_{ij}$ randomness.\cite{9} They showed that
the spatial variation in $J_{ij}$ might be irrelevant at long
wavelengths. For this reason, less attention was paid on this type
of disorder. However, the case that $J_{ij}$ obeys a discrete
probability distribution has still been intensively studied. A
simple choice of $P(J_{ij})$ is $P(J_{ij}) = p\delta(J_{ij}-J_0) +
(1-p)\delta(J_{ij}),$\cite{10,11,12} which means each bond might be
vacant with probability $1-p$. This model is often referred to as
the bond diluted model.

In fact, the continuous bond randomness case that $J_{ij}$ obeys a
continuous probability distribution should not be neglected.
Korshunov argued that the continuous bond randomness of the coupling
coefficient can greatly change the critical behavior of the BKT
transition, as long as the randomness is strong enough.\cite{13}
Recently, this point has also been discussed on the basis of the six-state clock
model.\cite{14} These works motivate us to study the 2D
XY model with continuous bond randomness.

It is plausible that concrete forms of different probability distributions
of the coupling may influence on the BKT transition differently. In
this paper we try to make sure whether it is true or not. We consider two
kinds of bond randomness, for which the distribution function,
$P(J_{ij})$, obeys the Gaussian distribution or uniform
distribution.

The 2D XY model under present consideration is defined on a square
lattice. For the first case, $P(J_{ij})$ is given by
\begin{align}
    P(J_{ij}) = \frac{1}{\sqrt{2\pi\sigma^2}}\exp\Bigl[-\frac{(J_{ij}-J_0)^2}{2\sigma^2}\Bigr],
\end{align}
where $J_0=1$ is the mean value of the coupling coefficients and $\sigma^2$ is
the variance. For the second case, $J_{ij}$ distributes uniformly in
the region $[J_0-d, J_0+d]$,
\begin{equation}
    P(J_{ij}) = \frac{1}{2d}=const.
\end{equation}
And variance of the uniform distribution is determined as
\begin{equation}
\sigma^2 = \braket{(J_{ij}-J_0)^2}= \frac{d^2}{3}.
\end{equation}

\begin{figure}[tb]
    \centering
    \includegraphics[width=0.45\textwidth]{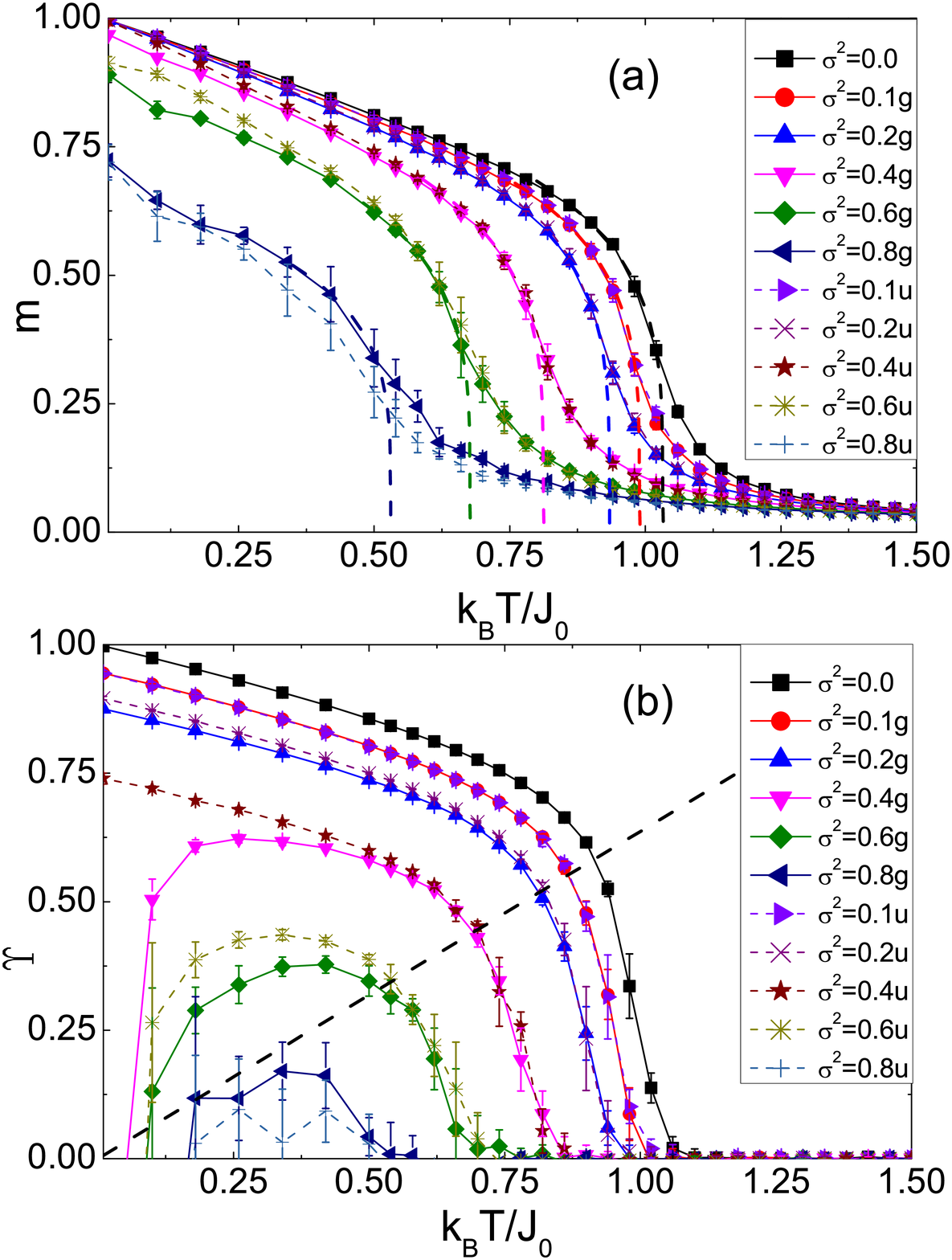}
    \caption{Finite-size magnetization (a) and helicity modulus (b) versus temperature for
    the random bond XY model with Gaussian distribution (marked as g) and
    uniform distribution (marked as u) of couplings. The dash
    lines in (b) and (a) plot Eq.(14) and  fitting curves for Gaussian-type
    magnetization based on Eq.(13)  respectively. The error bars are obtained
    using the standard deviations of numerical results.}
   \label{fig1}
\end{figure}

\begin{figure}[tb]
    \centering
    \includegraphics[width=0.45\textwidth]{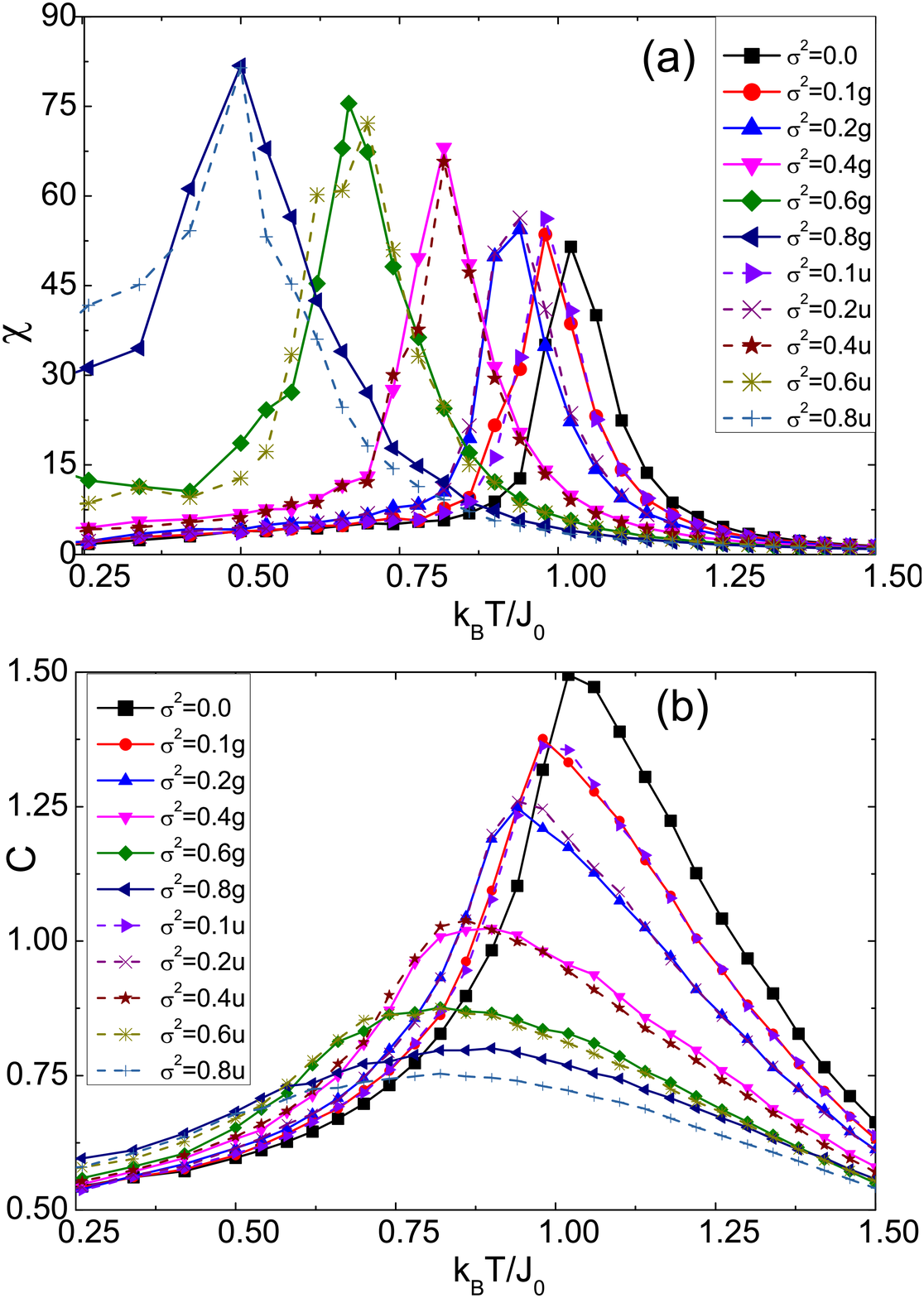}
      \caption{Susceptibility  (a), specific heat (b), versus temperature for
    the random bond XY model with Gaussian distribution (marked as g) and
    uniform distribution (marked as u) of couplings, respectively.}
   \label{fig2}
\end{figure}

We will characterize the BKT transition by investigating the
thermodynamics of the random bond 2D XY model. The thermodynamic
quantities calculated in the following include the finite-size
magnetization,
\begin{equation}
 m =\braket{|\textbf{\emph{M}}|},
\end{equation}
the susceptibility and the specific heat,
\begin{align}
\chi &= \frac{N}{k_B T} \Bigl(\braket{\bf{\emph{M}^2}} - \braket{\bf{\emph{M}}}^2\Bigr),\\
C &= \frac{N}{k_B T^2} \Bigl(\braket{{H^\prime}^2} -
\braket{H^\prime}^2\Bigr),
\end{align}
where
\begin{align}
 \textbf{\emph{M}} = (M_x,M_y) &= \Bigl(\frac{1}{N}\sum_i \sin\theta_i,
 \frac{1}{N}\sum_i \cos\theta_i\Bigr),\\
 H^\prime &= -\frac{1}{N}\sum_{\braket{ij}}J_{ij}{\bf S}_i\cdot {\bf S}_j,
\end{align}
$N$ is the number of sites and $\braket{}$ denotes thermodynamic
average. To calculate $\chi$ and $C$, we record $\textbf{\emph{M}}$ and
$H^\prime$ after each monte carlo step and exploit their
fluctuations.\cite{15}

In addition, the helicity modulus is a useful parameter to study the
BKT transition for its characteristic feature of  discontinuous
universal jump to zero at critical temperature.\cite{5} The
helicity modulus takes the form\cite{16}
\begin{align}
\braket{\Upsilon} = \braket{e} - \frac{N}{k_B T}\braket{s^2},
\end{align}
where
\begin{align}
e &\equiv \frac{1}{N}\sum_{\braket{ij}_x} \cos(\theta_i-\theta_j), \\
s &\equiv \frac{1}{N}\sum_{\braket{ij}_x} \sin(\theta_i-\theta_j),
\end{align}
The notation $\braket{ij}_x$ means the sum is over all links in one
direction only.

All the thermodynamic quantities are calculated numerically by
employing the standard Metropolis Monte Carlo method with
periodic boundary condition.\cite{17} The square lattice includes $N=64\times
64$ sites and we perform $10^6$ Monte Carlo steps to produce each
numerical result.

In Fig.1 and Fig.2, we plot thermodynamic quantities as functions of
temperature for both a Gaussian distribution and uniform distribution
cases. For the two different distributions, all four
thermodynamic quantities match pretty well when their variances
$\sigma^2$ take the same value, especially at high temperatures. Our
results suggest that different probability distributions of
couplings bring about similar effect on properties of the random
bond 2D XY model. In addition, it seems that the variance is to some
extent a good parameter to evaluate the strength of the disorder of
the random system. However, the matching  between the two distribution cases
becomes worse if the randomness tends to be extremely strong, for
example, $\sigma^2\geq0.6$. A possible reason is that thermodynamic behaviors
become dependent on the concrete random distribution of coupling coefficients
in the strong random limit. The BKT scenario might be invalid
when the disorder is extremely strong.\cite{14}

\begin{figure}[tb]
    \centering
    \includegraphics[width=0.45\textwidth]{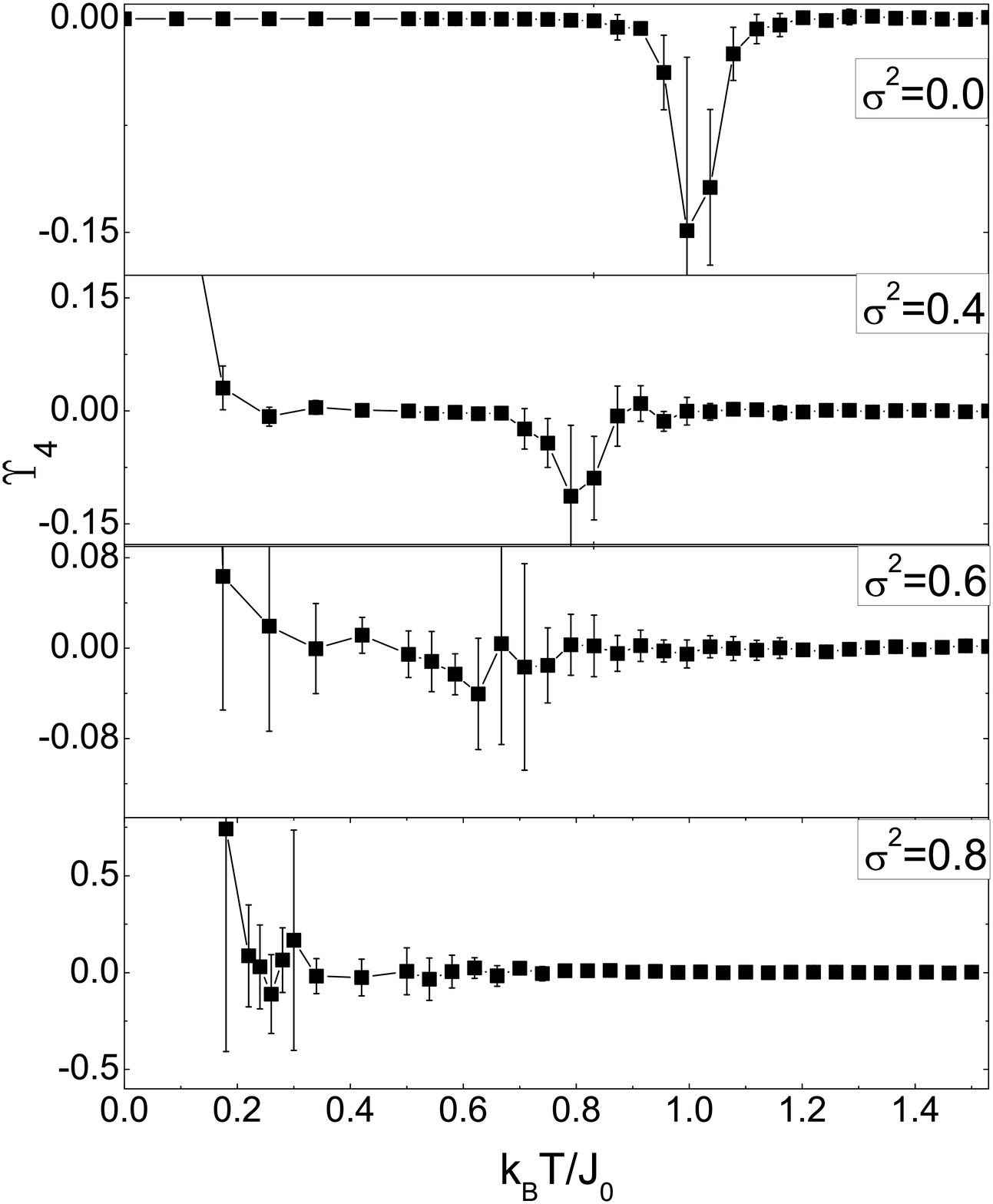}
    \caption{Fourth-order helicity modulus versus temperature for random
    bond XY model with Gaussian distribution of couplings.}
   \label{fig3}
\end{figure}

To proceed, we attempt to locate the BKT transition temperature
$T_c$. Usually, $T_c$ can be determined by the finite-size
magnetization and the helicity modulus in Fig.1, while in Fig.2 the
peak of susceptibility is not used to locate $T_c$ for lack of
accuracy,\cite{14,18} and the peak of specific heat occurs a certain
percent above $T_c$.\cite{19} The universal critical behavior of
finite size induced magnetization,\cite{4}
\begin{equation}
 m(T\to T^{\ast}) \sim (T_c-T)^{0.23},
\end{equation}
can be used to estimate the critical temperature by curve fitting,\cite{17}
where $T^{\ast}$ is a temperature near which the power law behavior
$(T_c-T)^{0.23}$ holds best. Fig.1(a) shows the fitting of magnetization.
For $\sigma^2=0.0$, $0.1$, $0.2$, $0.4$, and $0.6$, the estimated
value of $k_B T_c/J_0$ is $1.03, 0.99, 0.93, 0.81$, and $0.68$,
respectively.

The renormalization theory predicts the helicity modulus jumps from
$(2/{\pi})k_B T_{c}$ to zero in the thermodynamic limit.\cite{2}
It has been proved that this characteristic also exists in a bond diluted model,\cite{18,20}
which indicates it can be applied in our random
bond model. Therefore, $T_c$ can be estimated from the intersection
of $\Upsilon(T)$ and the straight line,\cite{18}
\begin{equation}
 \Upsilon=\frac{2}{\pi}k_{B}T .
\end{equation}
Estimating from intersecting points of curves and the straight line in Fig.1(b), we obtain that $k_B
T_c/J_0\approx0.92$, $0.87$, $0.81$, $0.70$, and $0.52$ for
$\sigma^2=0.0$, $0.1$, $0.2$, $0.4$, and $0.6$ respectively.

There is comparable difference between the BKT transition
temperatures determined by the two methods. It might be owing to the
finite size effect, since both methods to locate $T_c$ are
size-dependent but their size-dependence are different. In addition,
the obtained results of the helicity modulus lose accuracy when the
disorder turns strong, which can be seen in Fig.1(b) where the error
becomes quite large at $\sigma^2=0.6$ and $0.8$. Nevertheless, the
tendency that $T_c$ decreases with increasing $\sigma^2$ can be
confirmed, as shown in Figs.2(a) and 2(b).

For $\sigma^2=0.8$, curve fitting of the magnetization gives
$k_B T_c/J_0 =0.53$, while no appropriate intersection point can be found in
the helicity modulus in Fig.1(b). Thus a question arises that
what happens in the strong randomness limit, for example, at
$\sigma^2\geq0.8$. A possible answer to this question is that the
BKT transition is already destroyed by disorder in this case.\cite{14}

To further verify  the existence of the BKT transition we calculate
fourth-order helicity modulus $\Upsilon_4$ as suggested in
Ref.[16], which proved that the negative value of
$\Upsilon_4$ guarantees the discontinuous jump at $T_c$, and thus
guarantees the BKT transition. The fourth-order helicity
modulus can be expressed as
\begin{align}
 \braket{\Upsilon_4} =& -\frac{4}{N}\braket{\Upsilon} +
   3\biggl[\frac{\braket{e}}{N} - \frac{1}{k_B T}
   \braket{\bigl(\Upsilon-\braket{\Upsilon}\bigr)^2}\biggr] \nonumber\\
  &+\frac{2N^2}{T^3}\braket{s^4}.
\end{align}

The obtained result is shown in Fig.3. There exists a
trough around $T_c$ in $\Upsilon_4$. When $\sigma^2$ increases, the
depth of the trough may decrease but the negative of $\Upsilon_4$ at
$T_c$ is still quite clear for $\sigma^2=0.4$. For $\sigma^2=0.6$,
we can still identify that $\Upsilon_4$ is negative near $k_B T_c/J_0\approx
0.52$, despite the error in the data. However, as $\sigma^2$ reaches
$0.8$, the trough is overshadowed by noise and no discontinuous jump
of helicity modulus can be confirmed, which implies that the
BKT-type transition disappears.

In summary, thermodynamic properties of two dimensional random-bond
XY model on a square lattice are studied using Monte Carlo
simulations. The randomness may arise from disorder in real systems.
Two kinds of random-bond models are considered, with the probability
distributions of the coupling coefficient obeying the Gaussian
distribution and the uniform distribution, respectively. We show
that thermodynamic quantities of the two models are in good
agreement as long as their variances take the same value. Thus the
variance can be taken as a characteristic parameter to evaluate the
strength of the randomness. Moreover, it is shown that the BKT
transition temperature is suppressed as the randomness becomes
stronger. Furthermore, the BKT transition could even be destroyed in
the strongly disordered cases.

This work is supported by the National Natural Science Foundation of
China (Grant No. 11074021) and the Fundamental Research Funds for
the Central Universities of China.


\begin{thebibliography}{99}

\bibitem{1} Berezinskii V L 1971 {\it Sov. Phys. JETP} {\bf 32} 493

\bibitem{2} Kosterlitz J M and Thouless D J 1973 {\it J. Phys.} C {\bf 6} 1181

\bibitem{3} Kosterlitz J M 1974 {\it J. Phys.} C  {\bf 7} 1046

\bibitem{4} Bramwell S T and Holdsworth P C W 1993 {\it J. Phys. Condens. Matter} {\bf 5} L53 \\
            Bramwell S T and Holdsworth P C W 1994 {\it Phys. Rev.} B {\bf 49} 8811

\bibitem{5} Nelson D R and Kosterlitz J M 1977 {\it Phys. Rev. Lett} {\bf 39} 1201

\bibitem{6} Bishop D J and Reppy J D 1978 {\it Phys. Rev. Lett} {\bf 40} 1727

\bibitem{7} Resnick D J, Garland J C, Boyd J T, Shoemaker S and Newrock R S 1981 {\it Phys. Rev. Lett} {\bf 47} 1542

\bibitem{8} Trombettoni A, Smerzi A and Sodano P 2005 {\it New J. Phys} {\bf 7} 57

\bibitem{9} Rubinstein M, Shraiman S and Nelson D R 1983 {\it Phys. Rev.} B {\bf 27} 1800

\bibitem{10} Wu F Y 1982 {\it Rev. Mod. Phys} {\bf 54} 235

\bibitem{11} Surungan T and Okabe Y 2005 {\it Phys. Rev.} B {\bf 71} 184438

\bibitem{12} Zhu H X and Yan S L 2006 {\it Chin. Phys.} B {\bf 15} 3026

\bibitem{13}  Korshunov S E 1992 {\it Phys. Rev.} B {\bf 46} 6615

\bibitem{14} Wu R P H , Lo V C  and Huang H 2012 {\it J. Appl. Phys} {\bf 112} 063924

\bibitem{15} Newman M E J and Barkema G T 1999 {\it Monte Carlo Methods in Statistical Physics} (Oxford University Press) chap 1 pp 6-14

\bibitem{16} Minnhagen P and Kim B J 2003 {\it Phys. Rev.} B {\bf 67} 172509

\bibitem{17} Xu J and Gu Q  2012 {\it Phys. Rev.} A {\bf 85} 043608

\bibitem{18} Wysin G M, Pereira A R, Marques I A, Leonel S A and Coura P Z 2005
{\it Phys. Rev.} B {\bf 72} 094418

\bibitem{19} Gupta R and Baillie C F 1992 {\it Phys. Rev.} B {\bf 45 } 2883

\bibitem{20} Castro L M, Pires A S T  and Plascak J A 2002 {\it J Magn. Magn. Mater} {\bf 248} 62


\end{thebibliography}
\end{document}